\documentclass[named references]{kluwer} 
\newdisplay{guess}{Conjecture}
\usepackage{graphicx}
\begin{document}                                
\begin{article}
\begin{opening}         
\title{The Infrared Evolution of Sakurai's Object} 
\author{T. R. \surname{Geballe}}  
\institute{Gemini Observatory, Hilo HI 96720, U.S.A.}
\author{A. E. \surname{Evans}}
\author{B. \surname{Smalley}}
\author{V. H. \surname{Tyne}}
\institute{Physics Dept., Keele University, Keele, Staffordshire, ST5 5BG,
U.K.}
\author{S. P. S. \surname{Eyres}}
\institute{Astrophysics Research Institute, Birkenhead, CH41 1LD, U.K.}
\runningauthor{T. R. Geballe et al.}
\runningtitle{IR Observations of Sakurai's Object}

\begin{abstract}

Infrared spectroscopy and photometry have revealed the remarkable
evolution of Sakurai's Object from 1996 to the present.  A cooling,
carbon-rich photospheric spectrum was observable from 1996 to 1998.
Considerable changes occured in 1998 as the continuum reddened due to
absorption and emission by newly formed dust located outside the
photosphere. In addition, a strong and broad helium 1.083~$\mu$m P Cygni
line developed, signifying the acceleration of an outer envelope of
material to speeds as high as 1000~km~s$^{-1}$. At the same time the
photosphere of the central star remained quiescent. By 1999 the
photosphere was virtually completely obscured by the dust and the helium
emission line was the only detectable spectral feature remaining in the
1-5~$\mu$m band. In 2000 emission by dust has become even more dominant,
as the envelope continues to expand and cool and the helium line weakens.

\end{abstract}

\end{opening}           

\section{Introduction}  

This review is a phenomenological and chronological presentation of the IR
observations of Sakurai's Object. Its basis is the 1-5~$\mu$m spectroscopy
that the Keele-led team has acquired at the United Kingdom Infrared
Telescope (UKIRT) during the last four years. Those data are supplemented
in this paper with some of the other key infrared observations, both
spectroscopic and photometric. Over much of the last four years and over
much of the infrared region, emission by dust has dominated the spectrum
of Sakurai's Object. However, the emphasis here is on the description and
interpretation of the gaseous infrared features, as opposed to modelling
the dust emission, which is reviewed by Tyne (this volume).

\section{Overview}

Figure 1 presents in five panels an overview of the evolution of the
1-5~$\mu$m spectrum of Sakurai's Object from 1996 to 2000, as observed at
UKIRT with its facility spectrograph, CGS4 (Mountain et al. 1990). The
figure contains one spectrum from each of the five years. Each spectrum
consists of adjoined segments, with the resolving powers of the spectral
segments typically $\sim$200 to $\sim$2,000. Note that in the first two
years the data do not cover the entire 1-5~$\mu$m band. Some of the
spectra in Fig.~1 have been published in Eyres et al. (1998, 1999) and in
Tyne et al. (2000).

\begin{figure}
\centerline{\includegraphics[width=28pc]{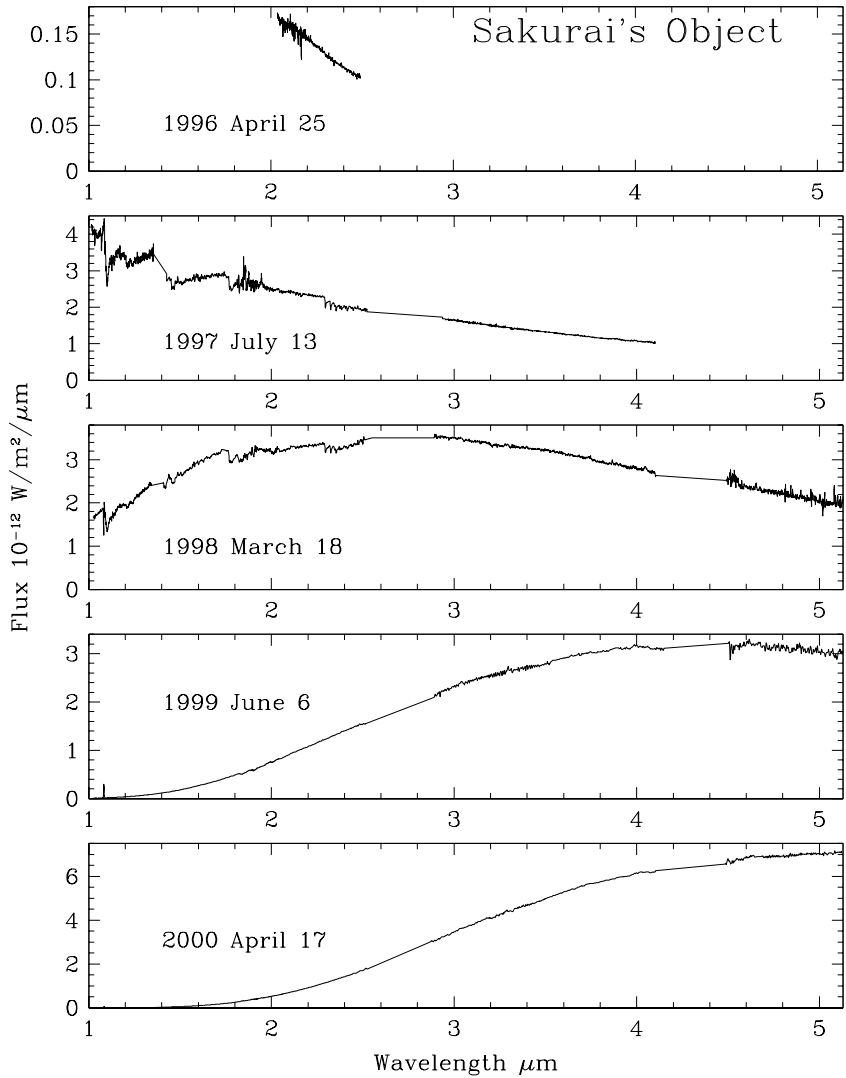}}
\caption{UKIRT/CGS4 spectra of Sakurai's Object from 1996 to 2000. The
spectra are interpolated in the gaps at 1.35-1.42~$\mu$m,
2.5-2.9~$\mu$m, and 4.1-4.5~$\mu$m.} \end{figure}

The 1996 spectrum reveals a photosphere in which atomic species are
dominant. In 1997 the photosphere had cooled to temperatures at which
molecules formed in abundance. During 1998 the photosphere became still
cooler and its spectrum increasingly reddened by newly forming dust. In
addition, during 1997-1998 continuum emission from the dust was becoming
dominant over the photospheric emission at thermal infrared wavelengths.
In 1999 the photospheric spectrum was no longer observable;  at the
shorter IR wavelengths it was completely obscured by dust absorption and
at longer wavelengths it was not only obscured, but also overwhelmed by
the dust emission. The dominance of the dust spectrum has has increased,
if anything, in 2000.

Even in the highly condensed presentation of the spectra in Fig.~1 it is
possible to pick out the most prominent narrow spectral features. The
strongest of these are the CN absorption bands near 1.1~$\mu$m,
1.2~$\mu$m, and 1.45~$\mu$m, which were present during 1997-1998, the
C$_{2}$ and CO bands observed during the same period near 1.8~$\mu$m and
2.3~$\mu$m, respectively, and the He~I 1.083~$\mu$m line, which has
undergone many remarkable changes since it first appeared in 1998 and has
been a key to understanding the events of the past three years.

\section{Year by Year Details}

\subsection{1996} 

From the limited infrared data available in 1996 (Fig.~1) the spectrum of
Sakurai's Object appears to be that of an F-type giant with abnormally
strong lines of atomic carbon. The slope of the IR spectrum and the lack
of CO at 2.3~$\mu$m and longward are consistent with this classification,
which generally agrees with the conclusions from the far more
extensive optical spectra and analysis by Asplund et al. (1997). A
portion of K-band spectrum is shown in Fig.~2. The large redshift of
Sakurai's Object, 113 km~s$^{-1}$ heliocentric (Duerbeck \& Benetti
1996), is readily apparent. The cluster of carbon lines near 2.12~$\mu$m
is much stronger than the same cluster in the sun, as are the probable CN
features near 2.143~$\mu$m. Duerbeck \& Benetti (2000) observed weak
Balmer lines in a spectrum obtained just one month prior to the K band
spectrum; based on their result one would not expect such a strong
Br~$\gamma$ (2.166~$\mu$m) absorption line as in Fig.~1. The line at
2.160~$\mu$m is unidentified.

\begin{figure}
\centerline{\includegraphics[width=28pc]{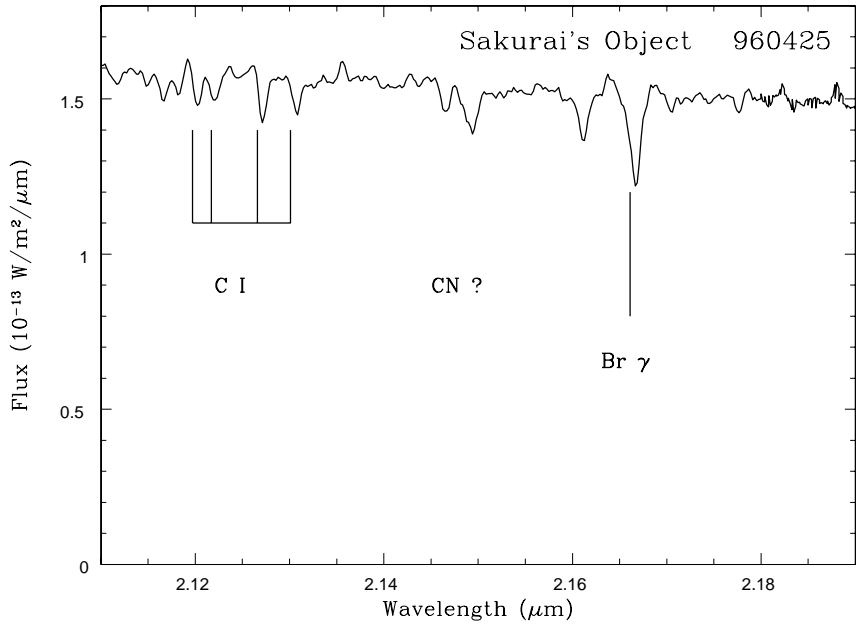}}
\caption{A portion of the K band spectrum of 1996. Vertical lines denote
laboratory wavelengths of identified lines.}
\end{figure}

\begin{figure}
\centerline{\includegraphics[width=28pc]{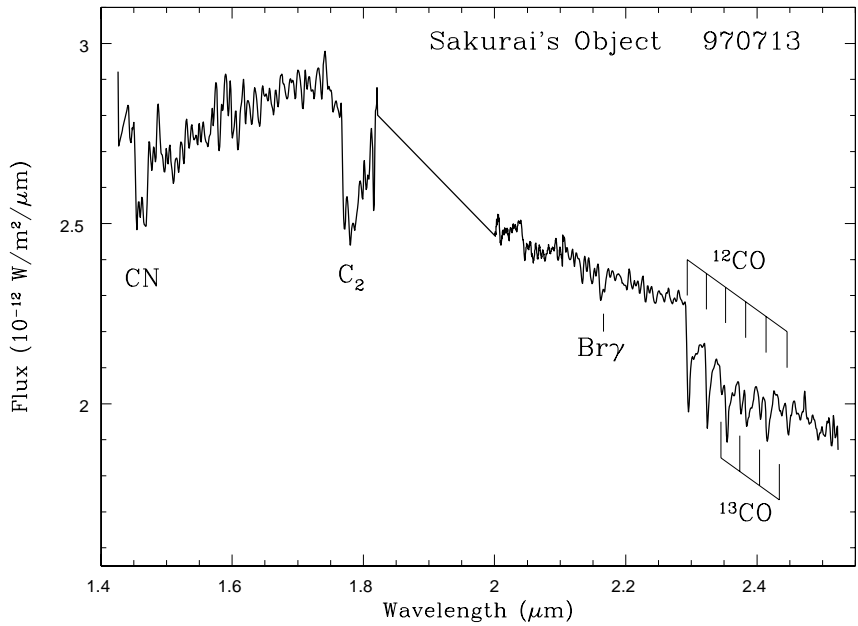}}
\caption{1.4-2.5~$\mu$m spectrum in 1997. Vertical lines for CO are
at the wavelengths of band heads.}
\end{figure}

\subsection{1997}

By 1997 the photosphere of the star had cooled considerably and molecules
had formed, although some atomic lines were still present. The strongest
spectral features were the CN bands longward of 1.1~$\mu$m and near
1.45~$\mu$m, the C$_{2}$ band near 1.8~$\mu$m, and the CO first overtone
band at 2.3~$\mu$m and longward. Most of these are shown in Fig.~3.  
Note that both $^{12}$CO and $^{13}$CO bands are present; and that at the
modest resolution of this spectrum (R$\sim$800) the $^{13}$CO band heads
are only 2--3 times weaker than the $^{12}$CO band heads. From this and a
high resolution spectrum obtained the following year (Fig.~6) it is
apparent that the $^{12}$C/$^{13}$C is very low, an inference consistent
with the results of Asplund et al. (1997) based on optical C$_{2}$ bands.
The Br~$\gamma$ absorption line was still present, although weaker than
in 1996. Lines of atomic strontium and carbon shortward of 1.1~$\mu$m
also were present (Eyres et al. 1998). Overall the continuum continued to
decrease to longer wavelengths in 1997, but more gradually than if it
arose only in the photosphere, whose temperature was estimated to be
5500~K by Pavlenko, Yakovina \& Duerbeck (2000).  This also was
apparent in the published photometry of Kamath \& Ashok (1999) and of
Kerber et al. (1999), and it led to suggestions that dust emission was
now present.  Eyres et al. fitted the excess seen in the UKIRT and ISO
data to graphitic dust at 680~K. Duerbeck et al. (2000) have pointed out
that free-free emission could contribute to the observed IR excess.

\subsection{1998}

\subsubsection{Massive dust formation}

In 1998 more substantial changes occurred. Kerber et al (1999) combined
ISO photometry with ground-based optical and near infrared photometry to
demonstrate that dust was forming on a much greater scale than previously.
In their Fig.~2 an obvious bump is seen in the spectral energy
distribution in 1998, peaking near 4~$\mu$m. No such feature existed in
1997. The infrared spectroscopic data from 1998 are consistent with this
change. The continuum no longer decreased monotonically with wavelength,
but showed a peak (in F$_{\lambda}$ just short of 3~$\mu$m (e.g., see
Fig.~1). The photospheric absorption bands due to CN, C$_{2}$, and CO were
still prominent, but the atomic features seen in previous years had
weakened, probably due to the continued cooling of the photosphere.

\begin{figure}
\centerline{\includegraphics[width=28pc]{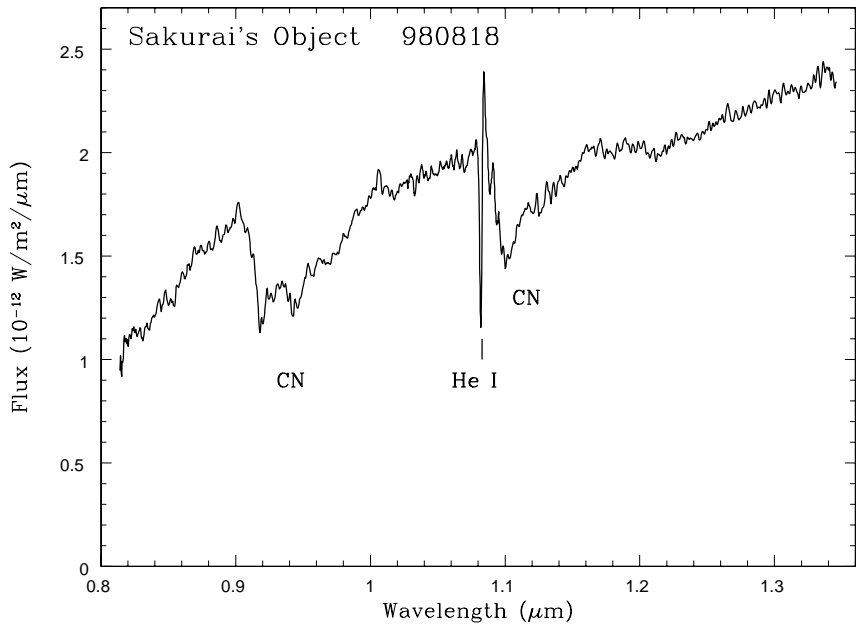}}
\caption{UKIRT/CGS4 I-J band spectrum in 1998}
\end{figure}

\begin{figure}
\centerline{\includegraphics[width=28pc]{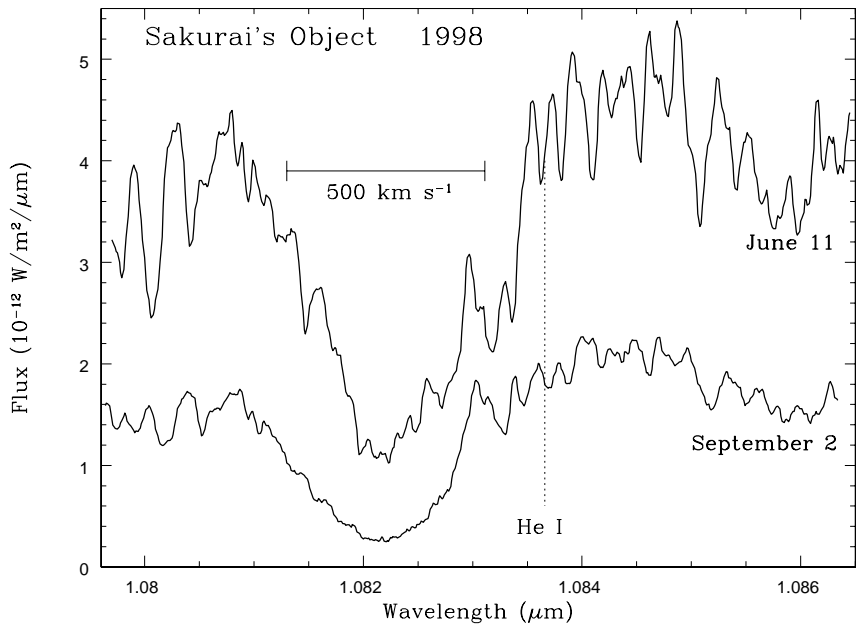}}
\caption{UKIRT/CGS4 high resolution spectra of the He~I
1.083~$\mu$m line in 1998. the dashed vertical line corresponds to the
wavelength of the helium line at the radial velocity of Sakurai's Object
as observed on June 11} 
\end{figure}

\subsubsection{The He I 1.083~$\mu$m line}

One prominent and new and spectral line appeared in 1998: the He~I line
at 1.083$\mu$m. It may be seen in Fig.~1, but is shown more clearly in
Figs.~4 and 5. Note that the spectrum in Fig.~4 extends shortward to
0.8~$\mu$m and reveals another strong CN absorption band near 0.9~$\mu$m.
The He line has a P Cygni profile, clearly signifying that mass loss is
occurring. The profile is seen in great detail in the high resolution
spectra (Fig.~5) from June and September of 1998. Each of the two spectra
in this figure shows an absorption trough extending about 800~km~s$^{-1}$
blueward of the stellar velocity. The redshifted emission extends about
400~km~s$^{-1}$ beyond the stellar velocity. Superposed on the profile
(and beyond it) are numerous narrower features, which are not identified.  
Some of these may be photospheric absorption lines, others might be
structure in the He line profile. The slight redshift of many of these
features from June to September is due to the change in the earth's
orbital velocity. Clearly the gas containing the observed helium line is
being accelerated outward from the star at high speed.

The 1.083~$\mu$m line is the allowed triplet 2P-2S transition of the
neutral helium. The lower (2$^{3}$S) level of the transition is
metastable and cannot radiate to or be radiatively excited directly from
the singlet (1S) ground state. The n=2 levels can be populated from below
by collisions or from higher energy states by radiative decay, for
example following recombination of He~II. In view of the lack of UV
radiation from the central star, photoionization of helium is unlikely.
He~II can be produced by collisional ionization. However, in 1998 the
helium line emission was weak compared to the absorption and and no
recombination lines of other elements were seen. Both of these suggest
that collisional ionization followed by recombination and radiative
cascading was not important. We tentatively conclude that in Sakurai's
Object the 2$^{3}$S level of He~I was populated mainly by collisions.

\begin{figure}
\centerline{\includegraphics[width=28pc]{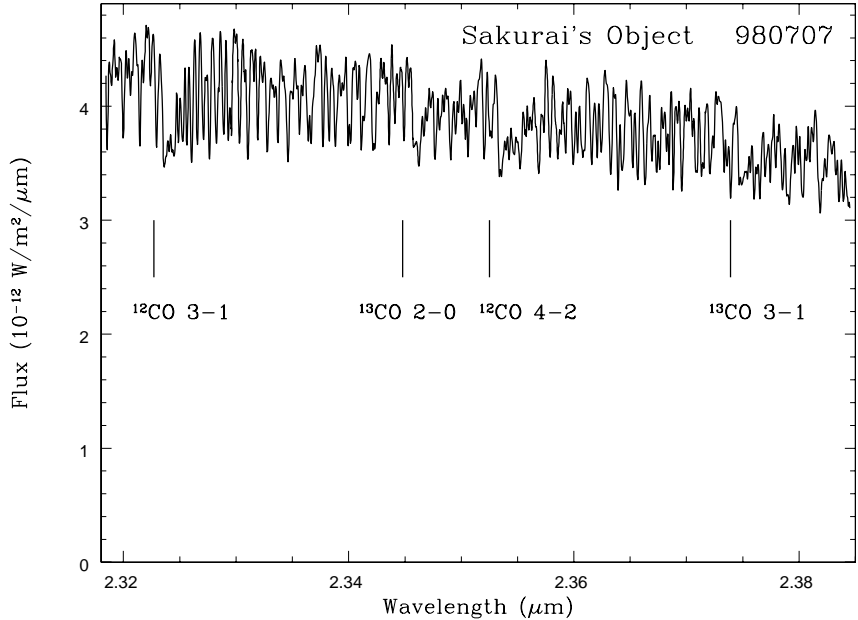}}
\caption{High resolution spectra of a portion of the CO first overtone
band. The laboratory wavelengths of band heads of $^{12}$CO and $^{13}$CO
are indicated and the heads are redshifted by roughly 0.002~$\mu$m.}
\end{figure}

It is important to note that the region in which the helium line profile
was formed in 1998 was completely detached from the photosphere of
Sakurai's Object.  This is evident from a comparison of Fig.~5 and
Fig.~6. The latter shows that a quiescent photosphere was present at same
time that the high velocity outflow was occuring. No sign of high speed
motions is evident in Fig.~6; the individual CO lines and the band heads
are sharp.  The photosphere was too hot for the dust to form there.
Presumably it was the formation of dust outside the photosphere and its
subsequent outward acceleration by radiation pressure that swept up and
heated the gas, as manifested by the 1.083~$\mu$m line. The broad
blueshifted absorption trough must have been produced by gas along the
line of sight to the continuum source, which at that time was probably
still mainly the photosphere of Sakurai's Object. The redshifted emission
likely arose in regions outside of this line of sight, presumably from
helium atoms in which the 2$^{3}$P (upper) level of the transition was
being populated, either by absorption of 1.083~$\mu$m photons by atoms in
the 2$^{3}$S level or by collisions. The great width of the absorption
component suggests that in 1998 the absorption was occuring very close to
where the continuum was formed and/or that a wide range of outflow speeds
were present along the line of sight to the continuum source. At later
times, when the ejected material becomes further detached from the
continuum source and possesses a narrower range of speeds, one would
expect the absorption profile to narrow and the emission to broaden.  
Close examination of Fig.~5 suggests that this effect was already being
seen in the sumer of 1998.

Toward the end of 1998 the visual--near IR brightness of Sakurai's Object
went into a steep decline. This is evidenced by the different continuum
levels of the two spectra in Fig.~5, from June and early September.
Spectra by Lynch et al. (this volume, see their Fig.~3) obtained at the
end of September show continued and more rapid change, with the flux at
1~$\mu$m reduced by nearly an order of magnitude compared to early
September. The profile of the He 1.083~$\mu$m line also was changing
radically at this time. Not only was the continuum decreasing but also
the broad absorption component was weakening. However, the emission
feature was roughly maintaining its strength. Thus in late September the
emission equivalent width had become much stronger than the absorption
equivalent width (Lynch et al, their Fig.~4).  The opposite was true four
weeks earlier (Fig.~5).

\subsection{1999 and 2000}

In April 1999, when spectra of Sakurai's Object were next obtained at
UKIRT (Tyne et al. 2000), the differences from the previous year's
spectra were profound. In the 1-2.5~$\mu$m region, where previously a
largely photospheric spectrum was observed, the flux was reduced, by
large factors at the shorter wavelengths, and a steeply rising continuum
was present (Fig.~7). Overall the peak of the spectrum had shifted to
$\sim$4.5~$\mu$m in 1999, corresponding to a mean dust temperature of
about 700 K, a cooling of roughly 300K in one year. By 2000 the peak had
shifted longward of 5~$\mu$m (Fig.~1), indicating further cooling of the
bulk of the dust.  The only possibly photospheric feature still observed
was a heavily obscured and veiled C$_{2}$ band near 1.78~$\mu$m (Fig.~7);
this band disappeared from view later in the year. The absorption
component of the He~I 1.083~$\mu$m line essentially had disappeared (it
is not apparent in UKIRT spectra from 1999, but a weak absorption is seen
in the July 1999 spectrum of Lynch et al., this volume).  The greatly
decreased 1~$\mu$m continuum and continued presence of helium line
emission suggests that the upper level of the He~I transition was being
excited mainly by collisions at that time, rather than by absorption of
1.083~$\mu$m radiation by atoms in the 2$^{3}$S state.

\begin{figure}
\centerline{\includegraphics[width=28pc]{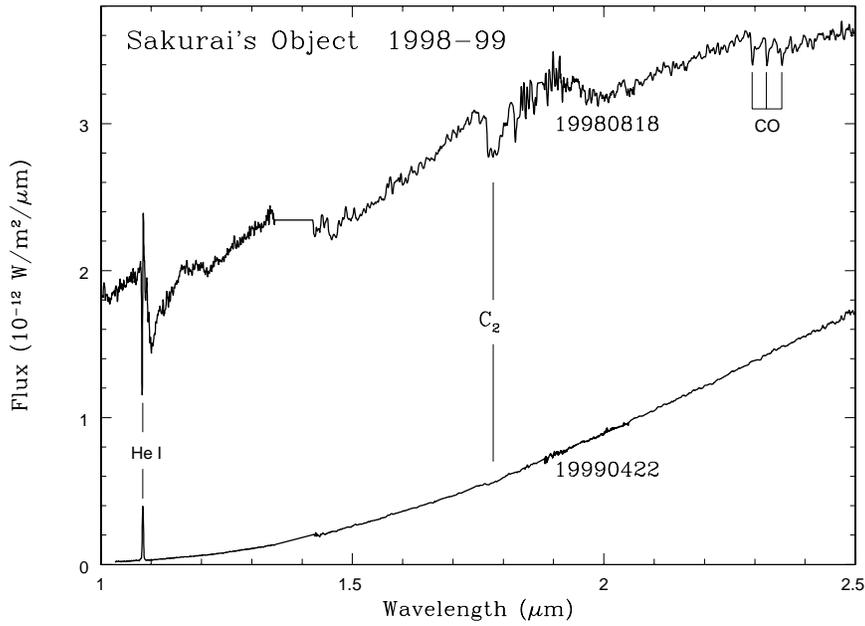}}
\caption{Comparison of 1-2.5~$\mu$m spectra in late 1998 and early 1999.} 
\end{figure}

\begin{figure}
\centerline{\includegraphics[width=28pc]{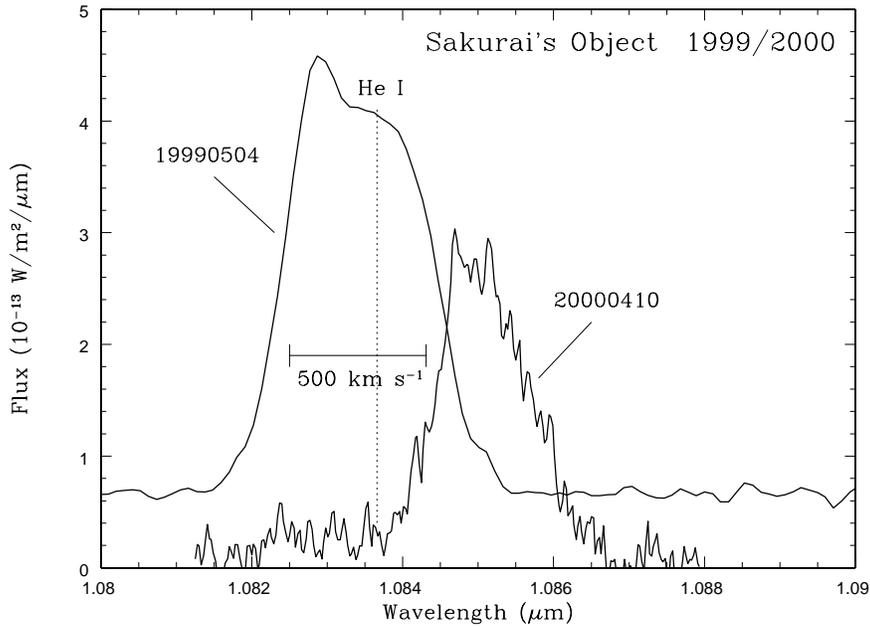}}
\caption{Helium line profiles in 1999 and 2000. The resolution of the
1999 spectrum is 60~km~s$^{-1}$; that of the 2000 spectrum, which has
been slightly smoothed, is 19~km~s$^{-1}$. The small fluctuations in the
latter are due to noise.}
\end{figure}   

Figure 8 shows representative helium line profiles in 1999 and 2000.
Compared to 1998 (e.g., Fig.~5) the 1999 profile was remarkably altered,
with absorption all but vanished and the center of the emission close to
the stellar velocity. A weak but blueshifted emission component was
present $\sim$1500~km~s$^{-1}$ from the stellar velocity. The gas
associated with this component was moving considerably more rapidly than
any gas observed previously. No similar red shoulder was seen. However,
the most highly redshifted gas should be situated directly behind the
star; if so its line emission would be obscured from view.

The helium line emission that still remains in 2000 is almost entirely at
positive velocities, and thus originates mostly in the rear half of the
expanding envelope. Line emission at blueshifted velocities has weakened
considerably since 1999, indicating that the rate of collisional
excitation of the n=2 levels of He I in the front of the envelope has
dropped. There appears to be no {\it a priori} reason this behavior of
the line profile, which naively would have been expected to remain
roughly centered on the stellar velocity as it faded. The behavior
suggests a non spherically symmetric geometry for the hot gas surrounding
Sakurai's Object or a late and asymmetric episode of ejection or
excitation.

\section{Future IR Observations}

The born again star at the center of Sakurai's Object is now obscured
from view, even at infrared wavelengths. Moreover, the peak of dust
continuum has now shifted longward of 5~$\mu$m, beyond where most
previous infrared measurements were concentrated. The only gaseous
spectral feature remaining in 2000, the He I 1.083~$\mu$m line, is fading
rapidly and may well be undetectable by 2001. Until the veil of dust
lifts and the further evolution of the the central star can be followed
directly, the most important observational needs are infrared photometry
across all ground-based windows and IR spectroscopy covering the 10 and
20~$\mu$m windows, in order to witness the further evolution of the dust
envelope.

\acknowledgements

We thank the staff of UKIRT for its assistance during our ongoing
observing campaign. The United Kingdom Infrared Telescope is operated by
the Joint Astronomy Centre on behalf of the U.K. Particle Physics and
Astronomy Research Council.

\end{article}
\end{document}